\documentclass[12pt,a4paper]{article}
\usepackage[latin1]{inputenc}
\usepackage{a4wide}
\usepackage{amsfonts}
\usepackage{amssymb}
\usepackage{amsmath}
\usepackage{bbm}
\usepackage{ifpdf}
\ifpdf
\usepackage[pdftex]{graphicx}
\usepackage[pdftex,unicode,implicit]{hyperref}
\hypersetup{%
  pdftitle    = {Supersymmetric non-Abelian  black holes and monopoles in Einstein-Yang-Mills sugras},
  pdfkeywords = {Supersymmetry, supergravity, gauge theories, black holes},
  pdfauthor   = {Mechthild Huebscher, Patrick Meessen, Tomas Ortin and Silvia Vaula},
  pdfcreator  = {pdf\LaTeXe\ with package \flqq hyperref\frqq},
  pdfproducer = {pdf\LaTeXe\ with package \flqq hyperref\frqq},
  pdfpagemode = None,  
  pdffitwindow= true,  
  unicode     = true,
  plainpages  = true,
  colorlinks  = true,  
  citecolor   = blue,
  urlcolor    = red,
  linkcolor   = black
}

\else
  \usepackage[dvips]{graphicx}
  \usepackage[unicode,implicit]{hyperref}

\fi
\makeatletter
\@addtoreset{equation}{section}
\makeatother

\makeindex
\begin{document}

\begin{flushright}
\small
IFT-UAM/CSIC-09-03\\
February $20^{\rm th}$ 2009\\
\normalsize
\end{flushright}

\begin{center}

\vspace{.7cm}

{\LARGE {\bf Supersymmetric non-Abelian  
black holes and\\[.5cm] monopoles in Einstein-Yang-Mills SUGRAS}}\footnote{ 
 \textsl{Contribution to the proceedings of the 4th Workshop RTN
  ''Constituents, Fundamental Forces and Symmetries of the Universe'' Varna
  (Bulgaria) 11-17 September 2008, to be published in Fortschritte der Physik
  - Progress of Physics }
}

\begin{center}

{\bf Mechthild H\"ubscher, Patrick Meessen, Tom\'as Ort\'{\i}n and Silvia Vaul\`a}

\vspace{.7cm}

{\em Instituto de F\'{\i}sica Te\'orica UAM/CSIC
Facultad de Ciencias C-XVI, \\
C.U. Cantoblanco, E-28049-Madrid, Spain\vskip 5pt}

{e-mail: {\tt \{Mechthild.Huebscher , Patrick.Meessen , Tomas.Ortin , Silvia.Vaula\}@uam.es}}

\end{center}

\vspace{1.5cm}

{\bf Abstract}

\begin{quotation}

 {\small    We discuss the construction of non-Abelian black holes and globally regular
   monopole solutions to $N=2$ $d=4$ EYM theories. 
   Special emphasis is put on how the attractor mechanism works for the 
   non-Abelian black holes.   
}

\end{quotation}

\end{center}

\newpage
\pagestyle{plain}

The class of supersymmetric solutions in ungauged $N=2$ $d=4$ supergravity
theories that can give rise to black hole spacetimes is completely
characterized \cite{Behrndt:1997ny}: the most general of such solutions is
parametrized by an even number of real functions that are harmonic on
$\mathbb{R}^{3}$ and to which we shall refer as {\em seed functions}.  Given
these seed functions there exists an algorithm to express all the physical
fields in terms of them.  Unsurprisingly, not every choice for the seed
functions leads to a regular black hole spacetime, but the necessary and
sufficient conditions that allow for a static, asymptotically flat,
spherically symmetric black hole are known (see {\em e.g.\/}
\cite{Bellorin:2006xr}). Apart from the absence of NUT charge and the
positivity of the ADM mass, the only remaining condition one has to impose is
the existence of a horizon. Having localized the horizon, the {\em near
  horizon geometry}, {\em i.e.\/} the geometry obtained by zooming in on the
horizon, is that of an $aDS_{2}\times S^{2}$ geometry and the entropy of the
black hole can then be read off easily by using the area law.
\par
The surprising outcome of the above story is that the entropy of the resulting
black hole depends only on the electric and magnetic charges of the various
Maxwell fields, and not on the asymptotic values of the scalar fields.  As
befitting uncharged scalar fields, they are constant on the horizon, but their
horizon values are also only given in terms of the electric and magnetic
charges of the Maxwell fields. This fact goes under the name of {\em attractor
  mechanism} \cite{Ferrara:1995ih} and is fundamental to the successful
matching of the entropy calculated in supergravity with the microscopical one
in string theory \cite{Strominger:1996sh}.
\par
These results have not been extended so far to black holes with non-Abelian
charges.  In order to obtain such solutions, the natural setup is
\emph{gauged} $N=2$ $d=4$ supergravity coupled to non-Abelian vector
supermultiplets.  {}For sake of simplicity we shan't introduce
hypermultiplets. Whence only isometries of the special-K\"ahler manifold
parametrized by the scalars in the vector multiplets are gauged and we shall
ignore possibility of turning on a Fayet-Iliopoulos term. We refer to these
models as $N=2$ $d=4$ Einstein-Yang-Mills (EYM) theories.  As in the Abelian
case, one can proceed to characterize the supersymmetric solutions in terms of
seed functions and select those that generate regular black hole solutions.
Unlike the Abelian case where the Bianchi and Maxwell equations imply that the
seed functions are harmonic on $\mathbb{R}^3$, however, the non-Abelian case
allows for less trivial field configurations.  We can in fact rephrase the
problem in terms of finding among the solutions to the Bogomol'nyi equations
for a given Yang-Mills-Higgs system, those that, when embedded into a suitable
EYM theory, give rise to regular non-Abelian black holes.  As an example we
will treat the embedding of unit-charge solutions to the ${\rm SO}(3)$ YMH
system into the $\overline{\mathbb{CP}}^3$ EYM model.
\par
Remarkably, we find that even in the non-Abelian case some kind of attractor
mechanism is at work.  In fact, we find well-defined hairy non-Abelian black
holes, with hair that is asymptotically invisible and does not contribute to
the properties of the horizon. In these hairy solutions the entropy of the
black hole depends only on the asymptotic \emph{color charges} (See {\em
  e.g.\/} Ref.~\cite{Volkov:1998cc} for more information on the definition of
color charge).  The scalars on the horizon depend also only on the color
charges, but are not constant over it, rather they form a hedgehog
configuration; this was to be expected as we are dealing with scalars
transforming in the adjoint of the gauge group.  The idea that the attractor
mechanism is expressible in terms of asymptotic charges is, however, not true
as will be seen by the construction of a {\em colored black hole}: for such
black holes, no asymptotic charges exist, yet the entropy of the black hole
receives contributions from the unit-charge YMH-configuration.
\par
{}As mentioned, we shall consider the $\overline{\mathbb{CP}}^3$ EYM model,
which is based on the special geometry of the $SU(1,3)/U(3)$ coset space.  The
model features one gravitational multiplet coupled to three vector
multiplets. The bosonic field content of this model consists of the metric, 4
gauge fields $A^{\Lambda}$ ($\Lambda = 0,1,2,3$) and 3 complex scalar fields
$Z^{i}$ ($i=1,\,2,\,3$). The fermionic field content consists in two gravitini
$\psi_{I\mu}$, ($I=1,\,2$) and six gaugini $\lambda^{Ii}$, two for each vector
multiplet, transforming as doublets under the ${\rm SU}(2)$ R-symmetry
group. We refrain from gauging the ${\rm U}(2)$ R-symmetry group and choose
an $SO(3)$-gauging, where 3 of the vector fields, $A^{i}$, constitute the
gauge fields, leaving $A^{0}$ as an Abelian field; the scalars and the
doublets of gaugini transform as a triplet under the $SO(3)$.
\par 
The bosonic action for the model is
{\small \begin{equation}
  \label{eq:DefAction}
  \int_{4}\sqrt{g}\left[ 
        \textstyle{1\over 2}R
   \ +\ \mathcal{G}_{i\bar{j}}\mathfrak{D}_{a}Z^{i}\mathfrak{D}^{a}\overline{Z}^{\bar{j}}
   \ -\ V(Z,\overline{Z})
   \ +\ \mathrm{Im}(\mathcal{N})_{\Lambda\Sigma} F^{\Lambda}_{ab}F^{\Sigma ab}
   \ -\ \mathrm{Re}(\mathcal{N})_{\Lambda\Sigma} F^{\Lambda}_{ab}\ (\star F)^{\Sigma ab}
  \right] ,
\end{equation}}
and the field strengths and the covariant derivatives are given by
\begin{equation}
  \label{eq:DefCovDer}
  F^{0} \ =\ dA^{0} \;\; ,\;\; 
  F^{i} \ =\ dA^{i} +\textstyle{g\over 2}\varepsilon_{jk}{}^{i}A^{j}\wedge A^{k}
  \;\; ,\;\; 
  \mathfrak{D}Z^{i} \ =\ dZ^{i} + g\ \varepsilon_{jk}{}^{i}A^{j}\ Z^{k} \; ,
\end{equation}
where $g$ is the coupling constant.  As the metric $\mathcal{G}$ is K\"ahler
it can be derived from a K\"ahler potential, $\mathcal{K}$, which for the
chosen model reads $e^{-\mathcal{K}} = 1-|Z|^{2}$.  Observe that the K\"ahler
potential imposes the constraint $0\leq |Z|^{2} \leq 1$, but it can be shown
that a BH solution automatically satisfies this bound \cite{Bellorin:2006xr}.
The explicit form of the complex matrix $\mathcal{N}$ and the potential $V$,
which by construction is positive semi-definite, can be written down, but as
they will not be used we shall refrain from doing it.
\par 
In order to find supersymmetric solutions it is convenient to identify first
the bosonic field configurations that admit at least one Killing spinor. This
is achieved by looking for solutions to the Killing Spinor Equations (KSEs)
\begin{equation}
\delta_{\epsilon}\psi_{I\, \mu} =\delta_{\epsilon}\lambda^{Ii} = 0\,,\label{KSE}
\end{equation}
whose explicit form can be found {\em e.g.\/} in Ref.~\cite{Huebscher:2007hj}.
Afterwards one imposes the equations of motion, thus sorting the
supersymmetric solutions out of the supersymmetric configurations. This task
is specially simple due to the existence of Killing Spinor Identities (KSIs)
relating the off-shell e.o.m.s of supersymmetric configurations.  As a
consequence one must explicitly solve just a limited number of equations of
motion, the others being implied by the KSIs.  In the case of interest, that
is when the metric admits a timelike Killing vector, it is enough to verify
that the Maxwell equations and the Bianchi identities of the vector fields are
satisfied.
\par
The static metric admitting a timelike Killing vector can be conveniently
written in the form
\begin{equation}
 ds^{2} \; =\; K^{-1}\ dt^{2} \ -\ K\ dx^mdx^m,\quad\quad\quad m=1,\,2,\,3\,.
\end{equation}
The function $K$ determines the whole metric and is 
expressed in terms of the seed function according to 
\begin{equation}
K=\langle\mathcal{R},\,\mathcal{I}\rangle \ =\ \mathcal{I}^{\Lambda}\mathcal{R}_{\Lambda}
     \ -\ \mathcal{I}_{\Lambda}\mathcal{R}^{\Lambda}\label{Kdef} 
\end{equation}
where $\mathcal{I}\equiv(\mathcal{I}^\Lambda,\,\mathcal{I}_\Lambda)$ is the
symplectic vector of real, seed functions, and $\langle\ ,\, \rangle$ is, as
indicated, a symplectic product.  The symplectic vector
$\mathcal{R}\equiv(\mathcal{R}^\Lambda,\,\mathcal{R}_\Lambda)$ is not
independent but is related to $\mathcal{I}$ via the model-dependent
stabilization equations.  {}For our chosen model the solution to the
stabilization equations read
\begin{equation}  
\mathcal{R}_\Lambda=-\textstyle{\frac12}\eta_{\Lambda\Sigma}\mathcal{I}^\Sigma;
\quad\quad\mathcal{R}^\Lambda=2\eta^{\Lambda\Sigma}\mathcal{I}_\Sigma \; .\label{stabil}
\end{equation}
The seed functions are defined in such a way that the scalar fields read
\begin{equation}
  Z^i=\frac{\mathcal{R}^i+2i\,\mathcal{I}^i}{\mathcal{R}^0+2i\,\mathcal{I}^0}\, ,
\end{equation}
while the expression for the field strengths provided by the KSEs is
\begin{equation}
F^{\Lambda}\ =\ -\textstyle{1\over\sqrt{2}}\ \mathfrak{D}(K\mathcal{R}^\Lambda\ dt)
             \ -\ \textstyle{1\over \sqrt{2}}\ K\star(dt\wedge\mathfrak{D}\mathcal{I}^\Lambda) \; .\label{F}
\end{equation}
Staticity imposes the following relation among the seed functions
\begin{equation}
  \langle\mathcal{I},\,\mathfrak{D}_m\mathcal{I}\rangle=0\,.\label{stat}
\end{equation}
From (\ref{F}) we can impose the Bianchi identities of the vector fields,
which turns out to be equivalent to
\begin{equation}
\mathfrak{D}_m\mathfrak{D}_m\mathcal{I}^\Lambda=0\label{BI} \; .
\end{equation}
The Maxwell/Yang-Mills equations, on the other hand give rise to 
\begin{equation}
 \mathfrak{D}_m\mathfrak{D}_m\mathcal{I}_\Lambda \ =\ 
   \textstyle{g^{2}\over 2}\left[ 
          \varepsilon_{\Lambda (\Sigma}{}^{\Gamma}\varepsilon_{\Delta )\Gamma}{}^{\Omega}
          \mathcal{I}^{\Sigma}\mathcal{I}^{\Delta}
     \right]\ \mathcal{I}_{\Omega} \; ,
 \label{MX}
\end{equation}
where only the $\varepsilon_{ij}{}^{k}$ are non-vanishing.  Note that
(\ref{MX}) ensures the integrability of the staticity condition (\ref{stat}).
\par
Equations (\ref{BI}) and (\ref{MX}) are in general not easy to solve.  The
best strategy is to start with a given $A^\Lambda$, and try to deduce
$\mathcal{I}^\Lambda$ by comparing the resulting field strength in
Eq.~(\ref{F}).  Doing this implies that Eq.~(\ref{BI}) is equivalent to the
Bogomol'nyi equation
\begin{equation}
 \mathfrak{D}_p\mathcal{I}^\Lambda=-\textstyle{1\over \sqrt{2}}\ \epsilon_{pmn}F^\Lambda_{mn}\, ,
\end{equation}
where one should observe that the integrability condition for the Bogomol'nyi
equation is nothing but Eq.~(\ref{BI}). Moreover, the $\mathcal{I}^{\Lambda}$s
that are not charged under the gauge group, {\em e.g.\/} in our case
$\mathcal{I}^{0}$, are as before harmonic functions on $\mathbb{R}^{3}$.
\par
In the ungauged case, {\em i.e.\/} when $g=0$, the 6 seed functions in the
spherically symmetric case are given by
\begin{equation}
  \label{eq:UngSeed}
  \mathcal{I}^{\Lambda}\; =\; h^{\Lambda}\ +\ \frac{p^{\Lambda}}{r} \hspace{.3cm},\hspace{.3cm}
  \mathcal{I}_{\Lambda}\; =\; h_{\Lambda}\ +\ \frac{q_{\Lambda}}{r} \; , 
\end{equation}
where $p^{\Lambda}$ and $q_{\Lambda}$ are the magnetic and electric charges of
the 4 Maxwell fields.  As mentioned above, given the seed functions we can
generate the complete expressions of the fields, and in this letter, we shall
be interested in static BH-like solutions. Then, staticity imposes the
constraint $h_{\Lambda}p^{\Lambda}-h^{\Lambda}q_{\Lambda}=0$ which is
equivalent to imposing the vanishing of a possible NUT charge. Here we shall
solve this constraint by putting
$\mathcal{I}_{\Lambda}=\mathcal{R}^{\Lambda}=0$, so that we will be dealing
with purely magnetic solutions only.  The supergravity fields can then be
expressed in terms of the seed functions as\footnote{ We shall omit the
  explicit expression for the Maxwell fields and refer the interested reader
  to Ref.~\cite{Huebscher:2007hj}.  }
\begin{equation}
  \label{eq:UngSol}
  \vec{Z} \; =\; \frac{\vec{\mathcal{I}}}{\mathcal{I}^{0}}
  \hspace{.2cm},\hspace{.2cm}
  2\ K \; =\; \left(\mathcal{I}\right)^{2} \ -\ \vec{\mathcal{I}}^{2} \; .
\end{equation}
Without loss of generality we can normalize the solution to asymptote to
ordinary Minkowski space by putting $(h^{0})^{2}= 2+\vec{h}^{2}$. The ADM mass
can be seen to be $2M=h^{0}p^{0}-\vec{h}\cdot\vec{p}$ and must be taken to be
positive.
\par
As one can see, there is a possible horizon located at $r=0$; the condition
for the solution to describe the geometry outside a regular black hole is then
equivalent to the statement that the geometry in the limit $r\rightarrow 0$,
is that of a well-defined $aDS_{2}\times S^{2}$ space, with the $S^{2}$ being
spacelike.  In case this is true, the area of the 2-sphere gives rise to the
black hole's entropy.  Of course, if we just calculate the limit and identify
the would-be area of the 2-sphere without bothering its well-definedness, we
find an expression for area/entropy which might be negative.  By an abuse of
language, then, we say that the condition for the existence of a horizon is
equivalent to the positivity of the entropy.
\par
Applying the above reasoning to our Abelian set-up then means that we have a
black hole if
\begin{equation}
  \label{eq:UngEntropy}
  2\ S_{bh} \; =\; (p^{0})^{2} \ -\ \vec{p}^{2}\ > \ 0\; .
\end{equation}
Observe that this entropy, conforming to the attractor mechanism, only depends
on the magnetic charges.  {}Furthermore, the expression for the scalars on the
horizon reads $\vec{Z} =\vec{p}/p^{0}$, which also fully conforms to the
attractor mechanism.
\par
The general class of supersymmetric solutions to $N=2$ $d=4$ SUGRA with
YM-couplings was obtained in Refs.~\cite{Huebscher:2007hj}.  {}For the example
at hand the expression of the fields in terms of the seed functions is still
given by Eq.~(\ref{eq:UngSol}), but the seed functions are no longer harmonic
functions on $\mathbb{R}^{3}$; instead, they have to satisfy
\begin{equation}
  \label{eq:Bogo}
  \vec{\partial}^{2}\mathcal{I}^{0} \ =\ 0 \hspace{.3cm},\hspace{.3cm}
  \mathfrak{D}_{i}\mathcal{I}^{l} \ =\ -\textstyle{1\over \sqrt{2}}\varepsilon_{ijk}\ \mathcal{F}^{l}_{jk} \; ,
\end{equation}
where the last equation is the Bogomol'nyi equation for the $SO(3)$
Yang-Mills-Higgs system on $\mathbb{R}^{3}$ determining the pair
$(A^{l}_{i},\mathcal{I}^{l})$.  The most famous spherically symmetric solution
to the Bogomol'nyi equation is of course the 't Hooft-Polyakov monopole, which
is characterized by the fact that it is completely regular.  This regularity
enables one to construct a globally regular, asymptotically flat supergravity
solution by taking $\mathcal{I}^{0}$ to be a suitable constant
\cite{Huebscher:2007hj}.\footnote{ One immediate question is: Can any monopole
  solution to the B.~equation be embedded into SUGRA theories as to give rise
  to a globally regular solution? The answer to this question is negative as a
  counterexample, to wit an $SU(4)$ monopole in the so-called
  $\mathbb{Q}$-magic model, can be constructed \cite{Huebscher:2007hj}.  }
\par
Of course, the Bogomol'nyi equation admits more spherically symmetric
solutions than just the 'tHP monopole \cite{Protogenov:1977tq}, but not all of
them have the desired properties to be used as seed solutions for constructing
black hole solutions. The question of what solutions to the Bogomol'nyi
equation can be used to build BH solutions was addressed in
Ref.~\cite{Meessen:2008kb}, and the conclusion is that apart from the 'tHP
monopole, there is one family of solutions giving rise to hairy black holes
with magnetic color charge $1/g$, and an isolated solution that is similar to
a colored black hole in that its asymptotic color charge vanishes.
\par  
We can study purely magnetic solutions by putting $\mathcal{I}_{\lambda}=0$,
which automatically solves Eqs.~(\ref{stat}) and (\ref{MX}), and then, writing
$\mathcal{I}^{0}=\sqrt{2}\left( h + p r^{-1}\right)$ for convenience, the
SUGRA fields for the hairy black holes are given by
\begin{equation}
  \label{eq:Effe}
  \vec{Z} \ =\ -\frac{\mu}{g}\frac{rH_{s}(r)}{p\ +\ hr }\ \vec{n} 
  \hspace{.2cm},\hspace{.2cm}
  K \ =\ (h+p/r)^{2} \ -\ \textstyle{\mu^{2}\over g^{2}}\ H_{s}^{2} \; ,
\end{equation}
where $\vec{n}$ is the outward-pointing unit vector, $\mu$ is a positive
parameter measuring the vacuum expectation value of the `Higgs' field
$g\vec{\mathcal{I}}$ and the function $H_{s}(r)$ is given by
\begin{equation}
  \label{eq:Effe2}
  H_{s}(r) \; =\; \coth\left( \mu r\ +\ s\right) \ -\ \left(\mu r\right)^{-1} 
  \hspace{1cm}\mbox{with}\;\; s\geq 0\; .
\end{equation}
The fundamental characteristic of the above function is that it is a
monotonically increasing function on the interval $(0,\infty )$ with
asymptotic values $H_{s}(r\rightarrow\infty )=1$.  {}For $s\neq 0$, the
function blows up as $r^{-1}$ around $r=0$, whereas when $s=0$ it behaves like
$H_{s}\sim r$ around $r=0$; in effect, the solution with $s=0$ corresponds to
the 't Hooft-Polyakov monopole.
\par
As in the ungauged case, we can normalize the solution to asymptote to
Minkowski space by putting $h =\sqrt{1+\mu^{2}g^{-2}}$, after which a small
calculation shows that the mass is given by $M= hp +\mu g^{-2}$, and therefore
imposes the constraint $hp>0$ in order for the mass to be positive.  The
globally regular solution is found by putting $s=0$ and $p=0$; in contrast,
for any $s\neq 0$ we can build a black hole whenever the `entropy is
positive', which imposes
\begin{equation}
  \label{eq:Effe3}
  S_{bh} \, =\, (p)^{2} - \textstyle{1\over g^{2}} \ >\ 0
  \hspace{.3cm}\mbox{and then we have}\hspace{.3cm} 
  \vec{Z}|_{hor} \ =\ -\textstyle{1\over gp}\ \vec{n}\; .
\end{equation}
Observe that, as in the ungauged case, the near horizon solution is determined
in terms of the asymptotic charges only, but that the scalars are not constant
over the horizon: they form a hedgehog configuration. This is, however, hardly
surprising as we are dealing with charged scalar fields on a 2-sphere.  The
r\^ole of the parameter $s\neq 0$ is, seeing that it cannot be expressed in
terms of asymptotic data such as the mass, that of a black hole hair and
exemplifies the known fact that the no-hair theorems are not valid in EYM
theories.  The fact that the entropy doesn't depend on the hair eiher, seems
fortunate when thinking about entropy calculations in string theory, but a
deeper string theoretical insight into the nature of the hair is, as far as
the authors are aware, lacking, but highly desirable.
\par
The hairy black holes constructed above, seem to indicate that some kind of
attractor mechanism is at work in the solutions, but as the last example
shows, the story is more complicated as there are solutions whose asymptotic
charges vanish: in analogy with the solutions found in EYM theories, we shall
call this example a {\em colored black hole}, and is given by the expression
\begin{equation}
  \label{eq:Colour1}
  \vec{Z} \ =\ -\frac{\vec{n}}{g(p+hr)(1+\lambda r^{2})} \;\; ,\;\;
  K \ =\ (h+p/r)^{2} \ -\ \left[gr\left( 1+\lambda r^{2}\right)\right]^{-2}
  \;\;\; (\lambda > 0)\; .
\end{equation}
Normalizing the metric as above fixes $|h|=1$ and the asymptotic mass of the
solution becomes $M=hp=|p|$, so that the asymptotic data are independent of
the parameter $\lambda$.  In order to construct a black hole, then, we need to
have a non-vanishing horizon, which is readily calculated and leads to the
same result as in Eq.~(\ref{eq:Effe3}); the resulting spacetime is best
interpreted as an extreme Reissner-Nordstrom black hole surrounded by a cloud
of ``glue'', whose fall-off is fast enough so as to not to leave an asymptotic
imprint but does contribute to the horizon-geometries.
\par
One might ask oneself about how generic these colored black holes are. At the
time of writing, we know of no colored black hole in supergravity not related
in one way or another to the one presented here. But we expect more,
especially as the colored configurations in non-supersymmetric EYM theories
are the only really non-Abelian solutions, the ones having asymptotic color
charge being embedding of Abelian solutions (see {\em e.g.\/}
Ref.~\cite{Volkov:1998cc}).
\par
The existence of this colored black hole implies that a non-Abelian version
of the attractor mechanism cannot be simply based on asymptotic charges,
making the problem none the easier. In this respect it is also worth pointing
out that the attractor mechanism also works for a large class of
non-supersymmetric solutions, but in that case little to no analytic solutions
of non-Abelian BHs are known.  Apart from the attractor, the main questions
for non-Abelian solutions to supergravity theories involve quantum corrections
in that one would like to know if they modify/constrain the hair and whether
they are compatible with the globally regular monopole solutions.  On the
stringy side, the fundamental question is how to build them out of fundamental
constituents and to see how and if they implement hair.


\section*{Acknowledgments}

This work was partially supported by the Spanish MEC grants FPA2006-00783 and
FPU AP2004-2574 (MH), a MEC Juan de la Cierva scholarship (SV), the CAM grant
HEPHACOS P-ESP-00346, by the EU RTN grant MRTN-CT-2004-005104, the Spanish
Consolider-Ingenio 2010 program CPAN CSD2007-00042 and by the {\em Fondo
  Social Europeo} by means of an I3P contract(PM).


\end{document}